\documentclass[twocolumn,showpacs,preprintnumbers,amsmath,amssymb]{revtex4}

\usepackage{graphicx}
\usepackage{dcolumn}
\usepackage{bm}

\begin{document}

\preprint{ }

\title{New Superconducting Phases in Field-Induced Organic Superconductor 
$\lambda$-(BETS)$_{2}$FeCl$_{4}$}

\author{M. Houzet$^{1}$, A. Buzdin$^{1}$, L. Bulaevskii$^{2}$, and M. 
Maley$^{2}$}
\affiliation{$^1$ CPMOH, Universit\'{e} bordeaux 1, F-33405 Talence Cedex, 
France \\
$^2$Los Alamos National Laboratory, Los Alamos, NM 87545}

\date{\today}

\begin{abstract}
We derive the parallel upper critical field, $H_{c2}$, as a function of the 
temperature $T$ in quasi-2D organic compound $\lambda$-(BETS)$_{2}$FeCl$_{4}$, 
accounting for the formation of the nonuniform LOFF state.  
To further check the 2D LOFF model 
we propose to 
study the $H_{c2}(T)$ curve at low $T$ in tilted fields, 
where the vortex state 
is described by the high Landau level functions characterized by the index 
$n$. 
We predict a cascade of first order transitions between vortex 
phases with different $n$, between phases with 
different types of the symmetry at given $n$ and the change of the 
superconducting transition from the second order to the first order as 
FeCl$_4$ ions are replaced partly by GaCl$_4$ ions. 
\end{abstract}


\maketitle

Very recently magnetic-field-induced superconductivity has been observed
in the quasi two-dimensional (2D) organic conductor $\lambda $-(BETS)$_{2}$
FeCl$_{4}$ \cite{Uji01}.  At zero magnetic field the antiferromagnetic
ordering of Fe$^{3+}$ moments in this compound gives rise to a metal-insulator
transition at temperatures around 8 K. 
A magnetic field above 10 T restores the metal phase. Further increase of the 
magnetic field induces the superconducting transition at $B\approx 17$ T. 
As it has been revealed in the very high magnetic field experiments \cite
{Balicas01}, the maximum critical temperature $T_{c}\approx 4$ K
is reached at $B_{0}\approx 33$ T and $T_{c}(B)$ drops to 2 K at $B=42$ T. 
Such an unusual behavior was interpreted in 
Refs.~\cite{Uji01,Balicas01} as 
a manifestation of the Jaccarino-Peter (JP) effect \cite{Jaccarino}, when
the exchange field of aligned Fe$^{3+}$ spins compensates the external
field in their combined effect on the electron spins. 
The JP compensation effect 
has been already proved to be responsible for the magnetic-field-induced
superconductivity in pseudo-ternary Eu-Sn molybdenum chalcogenides \cite
{Meul84} and in CePb$_{3}$ \cite{Lin85}. In contrast to $\lambda $-(BETS)$
_{2}$FeCl$_{4}$, in Eu-Sn molybdenum chalcogenides the superconducting
transition also exists in the absence of a magnetic field, and as the magnetic 
field increases, it first destroys superconductivity, but further restores it
in the range of fields 4 T$<B<$ 22 T at very low temperatures \cite{Meul84}. 
In CePb$_{3}$ in zero magnetic field the
heavy-fermion antiferromagnetic ground state was found \cite{Lin85}.
Similarly to $\lambda $-(BETS)$_{2}$FeCl$_{4}$, the magnetic field destroys
the antiferromagnetic ordering and induces superconductivity in fields higher
than 14 T. 

High-field superconductivity exists in all these compounds not only 
due to the JP effect, but also because of strong suppression of 
the orbital effect of the external field.  In CePb$_{3}$ and Eu-Sn
molybdenum chalcogenides the orbital effect is weak due to a very short 
superconducting coherence length in the presence of impurities.  In contrast, 
in the clean 
$\lambda $-(BETS)$_{2}$FeCl$_{4}$ superconductor strong quasi-2D 
anisotropy leads to negligible orbital effect for the field applied 
parallel to the layers.  Here the parallel upper critical field is
controlled by the spin effect only.  Then the temperature 
dependence of $H_{c2}$ may be well
described by the quasi-2D (Josephson-coupled-layers) BCS model 
\cite{Bulaevskii73}.  In this model at low temperatures $T<0.55~T_c$, 
the Larkin-Ovchinnikov-Fulde-Ferrell (LOFF) state is 
a correct superconducting phase \cite{FFLO,Bulaevskii74}.  
We note that previous attempts to identify unambiguously the LOFF state were 
unsuccessful because in 3D crystals orbital effect dominates over spin effect, 
while quasi-2D systems, like intercalated dichalcogenides, were not clean 
enough to allow for the LOFF state.  The very clean 2D organic crystal 
$\lambda $-(BETS)$_{2}$FeCl$_{4}$ with clear evidence for dominant spin 
effect is the most favorable system to 
detect the LOFF state unambiguously.   

In this Letter we use a 2D LOFF model for the clean superconductor 
$\lambda $-(BETS)$_{2}$FeCl$_{4}$ to calculate the  
$H_{c2}(T)$ curve for the orientation of the magnetic field parallel to the 
layers, accounting for the exchange field induced by Fe$^{3+}$ polarized 
magnetic moments.  We show that 
accounting for the 2D LOFF state improves fitting at low temperatures in 
comparison with the 3D model as was also 
discussed in Ref.~\cite{Balicas01}.  In addition to results 
\cite{Balicas01}, we 
calculate $H_{c2}$ for the magnetic fields tilted by angle 
$\theta$ with respect to 
the layers.  We show that a quasiperiodic-like angular dependence of 
$H_{c2}$ is 
a specific property of the vortex state, which originates from the 2D 
LOFF phase 
when the orbital effect, caused by the perpendicular 
component of the external field, is turned on as $\theta$ increases.  
This vortex structure is characterized 
by the discrete 
variable $n$, the Landau level index.  Changes of $n$ with $\theta$ lead
to a cascade of first order transitions of the vortex lattice.  
We argue also that partial replacement of FeCl$_4$ ions by nonmagnetic 
GaCl$_4$ 
ions should lead to the suppression of the LOFF state and to the first 
order normal-to-superconducting (N-S) transition. 

We consider first the $H_{c2}(T)$ curve for the parallel orientation of 
the external field. 
The total effective field acting on the electron spins is 
$H_{eff}=B-I\left\langle S\right\rangle/\mu$,  
where $\mu\approx \mu_B$ is the electron magnetic moment, and 
$I$ is the exchange integral for the interaction between the conduction 
electrons
and Fe$^{3+}$ magnetic moments.  The negative sign 
implies the JP compensation
effect.  In the whole ($B,T$)-region, where the field-induced
superconductivity exists in $\lambda $-(BETS)$_{2}$FeCl$_{4}$, the Fe$^{3+}$
moments are saturated, $\langle S\rangle =S_{0}$.  The maximum $T_c(H)$
corresponds to $H_{eff}=0$.  Hence, the 
effective field acting on spins is simply 
$H_{eff}=B-B_{0}$, where $B_0=IS_0/\mu$.
The upper critical field is given by the solution of the linearized 
equation for the superconducting order parameter $\Delta ({\bf r})$, 
where ${\bf r}$ $=(x,y)$ is the in-plane coordinate:  
\begin{equation}
\Delta ({\bf r})=\int K_{0}({\bf r}-{\bf r}^{\prime })\Delta ({\bf r}%
^{\prime })d{\bf r}^{\prime },
\end{equation}
For the nonuniform order parameter, 
$\Delta ({\bf r})= \Delta\exp(i{\bf q\cdot r})$, the kernel 
$K_0({\bf r}-{\bf r}^{\prime })$ takes the form \cite
{Bulaevskii73,Buzdin96}: 
\begin{equation}
K_0({\bf q})=2\pi \left| \lambda \right|T{\rm Re} \sum_{\omega >0}
\oint \frac{[2\pi 
|{\bf v}_F(l)|]^{-1}dl}
{\omega +i\mu H_{eff}+i{\bf v}_{F}(l){\bf q}/2}, \label{tc}
\end{equation}
where $\omega$ denotes the Matsubara's frequencies, $\lambda$ is the 
coupling 
parameter, line integral over the coordinate $l$ is along the 2D Fermi surface 
in the momentum space,  
${\bf v}_F(l)$ is the in-plane Fermi velocity.

As has been demonstrated by 
Larkin and Ovchinnikov and Fulde and Ferrell \cite{FFLO}, in magnetic 
fields above some 
critical value the modulated superconducting phase  with nonzero $q$ gives 
higher $T_c$ for singlet pairing.  
A uniform superconducting order parameter in the presence of uniform Zeeman 
splitting is unfavorable because electrons with opposite spins and momenta 
have different energies $\epsilon(k)\pm \mu H$, where $\epsilon(k)$ is the 
electron kinetic energy.  Meanwhile, the Cooper 
instability towards formation of superconducting pairs is strongest 
when the energies of electrons 
in a pair are close to each other.  Hence, pairing of 
electrons with momenta ${\bf k}\pm {\bf q}$ at $q\approx \mu H_{eff}/v_F$ 
and opposite spins is more favorable than at $q=0$, because energies 
$\epsilon({\bf k}+{\bf q})-\epsilon_F-\mu H_{eff}$ and 
$\epsilon({\bf k}-{\bf q})-\epsilon_F+\mu H_{eff}$ are 
closer.  Obviously, gain in the superconducting 
energy depends on the 
dimensionality of the electron motion.  In a quasi 1D system one can 
compensate 
Zeeman splitting almost completely, while in the 2D case compensation is only 
partial due to different orientations of the Fermi velocity, and in 3D 
crystals compensation is even less effective. 
Hence, 
$H_{c2}$ is very sensitive to the geometry of the in-plane Fermi surface 
(FS).  

\begin{figure}
\includegraphics[width=7cm]{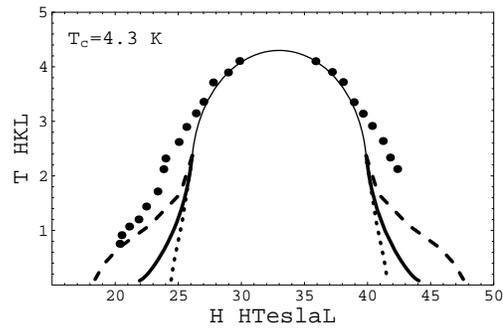}
\caption{\label{fig1} The dependence of the critical temperature on the magnetic field 
parallel to the layers in the presence of the Jaccarino-Peter effect. 
The thin solid curve is the transition from the normal metal to the uniform 
superconducting 
state. The dotted (thick) line is the transition to the LOFF state in 
the isotropic 
3D (2D) system and the dashed line is for the in-plane anisotropic 2D system. 
Dots are experimental data obtained by resistivity measurements [2]. }
\end{figure}

In the following we consider an isotropic in-plane electron spectrum and 
also the anisotropic one defined as 
\begin{equation}
\epsilon({\bf p})=W(\cos p_x+\Gamma\cos p_y), \ \ \epsilon_F=0, 
\label{sp}
\end{equation}
where the parameter $\Gamma\leq 1$ accounts for the in-plane anisotropy.  
$T_c(B)$ for such a nonuniform state is defined by the equation 
$K({\bf q},B,T)=1$. 
The wave-vector ${\bf q}$ at the transition from the normal state to the 
LOFF state is that which gives a 
maximal $H_{c2}$ at a given $T$.  It depends on the $v_F$ value, but $H_{c2}$ 
does not. 
The corresponding $H_{c2}(T)=B_{0}\pm H_{eff}$ curves for the 3D spherical 
FS, 2D circular FS 
and for the spectrum given by Eq.~(\ref{sp}) at $\Gamma=0.7$, 
calculated for $T_{c}= 4.3$ K, are shown in Fig.~\ref{fig1} together with 
the experimental data \cite{Balicas01}.  

The common part of the $B_{c2}(T)$ curves corresponding to the transition into 
the uniform state is presented by the thin solid line. 
To give insight into the importance of the in-plane anisotropy we 
calculate the critical value of $h=2\mu H_{eff}/\Delta_0$ at $T=0$ as a 
function of 
$\Gamma$.  Here $\Delta_0=1.76 T_c$ is the superconducting gap at $T=0$.  
The upper critical value of $h$ is given 
by the equation
\begin{equation}
\int_0^{\pi/2}du\frac{\ln\left|h^2-q_x^2(1-\Gamma^2\cos^2u)\right|}
{\sqrt{1-\Gamma^2\cos^2u}}=0. 
\end{equation}
At $\Gamma=0.7$ we obtain the upper critical value $h_{cr}=2.64$ 
corresponding to 
$\mu H_{eff}=1.32\Delta_0$.  For comparison, we 
obtain $h_{cr}=2$, corresponding to $\mu H_{eff}=\Delta_0$, in the case of 
a circular FS. In this approach $h_{cr}(\Gamma)$ 
increases without saturation as $\Gamma$ decreases.  

To check whether the LOFF
state is indeed realized in $\lambda $-(BETS)$_{2}$FeCl$_{4}$ a decisive
experimental test is needed.  Here we propose that such a critical experiment
may be the study of the dependence $H_{c2}(T)$ for
tilted-toward-the-plane field orientation.  The perpendicular component of the
magnetic field suppresses pairing by the orbital mechanism and leads to 
the formation of the vortex state on the already 
nonuniform LOFF background caused by the spin effect.  
The interplay between orbital and the spin effects
gives rise
to a very peculiar upper-critical-field behavior \cite{Bulaevskii74,Buzdin96} 
resulting from the solutions with higher Landau level functions for 
the superconducting order parameter in the vortex state.

To derive the $H_{c2}(T)$ curve at tilting angles $\theta$ we 
account for the effect of the vector potential describing the perpendicular 
component of the magnetic field, ${\bf B}_{\perp}$, 
in addition to the spin effect of $B_{\parallel}$ 
and of the exchange field.  
Now, in Eq.~(1) the kernel is given as 
\begin{equation}
K({\bf r},{\bf r}^{\prime })= 
K_{0}({\bf r}-{\bf r}^{\prime })\exp \left[
2ie/\hbar c\int_{{\bf r}^{\prime }}^{{\bf r}}{\bf A}({\bf s})d{\bf s}\right], 
\end{equation}
for an isotropic 2D FS and the in-plane vector-potential is 
${\bf A}=-({\bf B}_{\bot }\times {\bf r})/2,$ with $B_{\bot }=B\sin \theta $, 
\cite{Helfand}.
Following Refs.~\cite{Bulaevskii74,Buzdin96} general solutions of 
the integral equation for the order parameter are the Landau level functions:
\begin{equation}
\Delta_{n}({\bf r})=\exp(-in\varphi-\rho^{2}/2)\rho^{n},
\end{equation}
where $\rho =r\sqrt{eB_{\bot }/\hbar c}$ and $\varphi$ is the polar angle. 

\begin{figure}
\includegraphics[width=7cm]{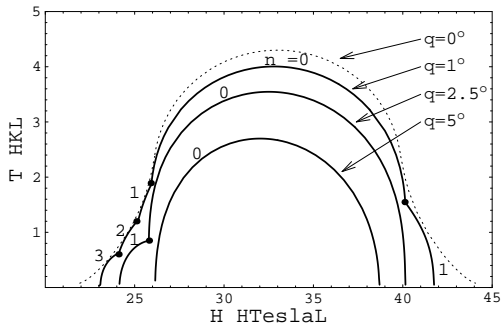}
\caption{\label{fig2}
The dependence of the upper critical  magnetic field 
on $T$ at different tilt angles $\theta$ for an isotropic 2D Fermi surface. 
Different parts of curve correspond to different Landau level 
functions characterized by the index $n$.   }
\end{figure}

When the orbital effect dominates, it is the zero Landau level function which 
gives the solution corresponding to the highest
magnetic critical field.  However, if the spin effect is strong 
and the electron mean 
free path is much bigger than the superconducting correlation length (i.e. 
in the clean limit), the higher Landau level 
functions for a superconducting nucleation  
give higher $H_{c2}$, because they provide 
rapid spatial variations of the superconductor order parameter needed to 
compensate Zeeman splitting of electrons in a singlet pair.  Discontinuous 
adjustments of the 
discrete variable $n$ to the change of the angle $\theta$ give rise to 
unusual quasi-oscillatory angular and temperature dependence of $H_{c2}$, see 
Refs.~\cite{Bulaevskii74,Buzdin96}.  The general expression
for $H_{c2}$ is given by the equation 
\begin{eqnarray}
&&\ln(T/T_{c}) =\pi T \times \label{selfB} \\
&&{\rm Re}
\sum_{\omega >0}\left\{ \int_{0}^{\infty }
\frac{(-1)^{n}L_{n}(x)\exp(-x/2)dx}{[(\omega +i\mu H_{eff})^{2}+
xv_{F}^{2}eB_{\bot }/4\hbar c]^{1/2}}-\frac{2}{\omega }\right\},
\nonumber
\end{eqnarray}
where $L_n(x)$ are the Laguerre polynomials.  
In Fig.~\ref{fig2} we show the expected behavior of $H_{c2}$ for different
tilting angles $\theta$ for the model with isotropic 2D FS. 
In these calculations we take the value 5 T for 
the orbital magnetic critical field at $T=0$ as was obtained for 
(BETS)$_2$GaCl$_4$ with similar conducting layers \cite{Uji01}.  
The behavior of the system in 
the quasi-1D limit may be calculated in the same way basing on the results of 
Ref.~\cite{Dupuis}. 

Next we discuss the vortex lattice originating from the LOFF state due to 
the orbital 
effect of ${\bf B}_{\perp}$.  Inside the superconducting phase, the
order parameter must form a lattice and without the spin effect 
it is a triangular Abrikosov vortex lattice made of $n=0$ Landau level 
functions.  In such a standard vortex lattice the number of zeros of the 
superconducting order parameter coincides with number of quantized vortices, 
i.e. there is one zero per unit cell.  
However,
for higher Landau level solutions the number of zeros exceeds number of 
quantized vortices determined by the component $B_{\perp}$ and only a 
fraction  
of the zeros coincides with the centers of quantized vortices. 
Respectively, zeros 
become inequivalent, they form a lattice determined by 
two length scales, the average distance between quantized vortices, 
$a=(\Phi_0/B_{\perp})^{1/2}$, and the LOFF period determined by the effective 
field, $\ell=\hbar v_F/\mu H_{eff}$.  This leads to 
a large variety of exotic vortex lattices as has been predicted in 
Refs.~\cite{Houzet 00,Klein 00}.  

\begin{figure}
\includegraphics[width=6cm]{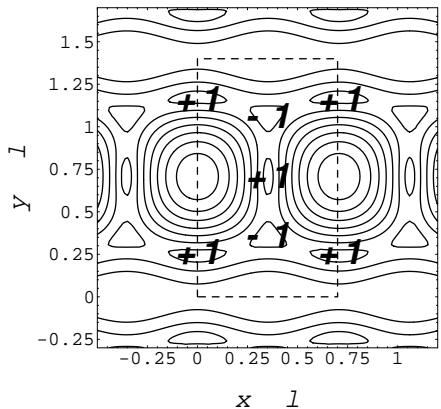}
\caption{\label{fig3}
Square asymmetric vortex lattice described by the Landau level functions 
$n=2$ formed in the presence of orbital and spin 
effects.  The profiles of the 
amplitude of the superconducting 
order parameter are shown.  The dashed line represents the unit cell with five 
zeros.  Their winding numbers, $\pm 1$, are shown. 
There is one flux quantum per unit cell. 
}
\end{figure}

To determine the equilibrium vortex structure 
we minimize the free energy.  In Ref.~\cite{Houzet 00} the
generalized Ginzburg-Landau (GL) functional has been derived to describe 
the LOFF 
state near the tricritical point, $H_{eff}^*=1.07~T_{c}/\mu, 
\ \ T^*=0.55~T_{c}$, for isotropic FS:
\begin{eqnarray}
&&\frac{{\cal F}}{N(0)T_{c}^{2}} =0.86~\frac{B-H_{eff}(T)}{H_{eff}^*}
|\Delta|^{2}+0.15~\frac{T-T^*}{T^*}\left|\Delta \right|^{4} \nonumber \\
&&+0.011~\left|\Delta\right| ^{6}
+3.0~\frac{T-T^*}{T^*}\xi_{0}^{2}|\tilde{ \nabla}\Delta|^{2}
+3.1~\xi_{0}^{4}|\tilde{\nabla }^{2}\Delta| ^{2}
 \label{energie}\\
&&+0.85~\xi_{0}^{2}\left\{\left|\Delta\right|^{2}|\tilde{\nabla}
\Delta |^{2}
+\frac{1}{8}\left[ ( \Delta^*\tilde{\nabla}
 \Delta )^{2}+(\Delta \tilde{\nabla}\Delta^*)^{2}
\right] \right\},  \nonumber 
\end{eqnarray}
where $N(0)$ is the DOS at 
the Fermi level, $\tilde{\nabla}=\nabla-(2ie/\hbar c){\bf A}$ and 
$\xi_{0}=\hbar v_{F}/2\pi T_{c}$ is the superconducting correlation length.  
Note that we need to include 
in the functional the second-order derivative term $\sim |\tilde{\nabla}
^{2}\Delta|^{2}$ due to the negative sign of the first-derivative term.  
Similarly, we add the sixth order term $\sim \left| \Delta \right|^{6}$ 
\cite{Buz-Kach}.  
Strictly speaking the derived functional is valid near the tricritical
point only, but we may expect that it describes qualitatively correctly  
(as the standard GL functional) the LOFF state at all 
temperatures.  Extending the analysis of the vortex lattice structure 
\cite{Abrikosov 57} to this new functional, one can calculate
the structures of possible new vortex lattices.  It turns out that the
N-S transition becomes of the first order in some
temperature and angle intervals \cite{Houzet 00} and the lattice 
structure exhibits zeros with positive and negative winding numbers 
$w$, $|w|\geq 1$, as well as strongly anisotropic 
structures.  
In Fig.~\ref{fig3} we present as an example the square asymmetric vortex lattice 
obtained numerically by use of the functional (\ref{energie}).  It 
corresponds to the Landau level function $n=2$ and have five zeros in the 
unit cell.  two of them has $w=-1$, while three have $w=+1$,  
so that there is one flux quantum per unit cell.  Note that zeros are 
positioned inequivalently and that different types of symmetry in their 
positions as well as different indecies $n$ lead to a 
variety of vortex lattices and first order phase transitions between them.  

Let us discuss now the anticipated behavior of the system 
$\lambda-({\rm BETS})_{2}({\rm FeCl}_{4})_{1-x}({\rm GaCl}_4)_x$ using 
information \cite{Ga} on $H_{c2}(T)$ at $x=1$.  Replacing the Fe$^{3+}$ ion 
randomly by nonmagnetic Ga$^{3+}$ reduces the average exchange field as 
$B_0(1-x)$. Additionally, a random exchange field leads to 
electron scattering with rate of order $\mu B_0x$. It causes suppression of 
the LOFF 
state at $x\approx 1/2$, when the scattering rate becomes comparable with 
$\Delta_0$. Hence, as $x$ increases, while remaining small, in Figs.~\ref{fig1}, \ref{fig2} 
the curves 
$T(H)$ scale as $(1-x)H$, while the thick and dashed 
lines approach the dotted line due to the suppression of the LOFF state.  At 
$x\approx 1/2$ we anticipate that 
due to strong exchange scattering the LOFF state vanishes, and the first 
order superconducting 
transition takes place near the dotted line, at 
$H_{eff}=\Delta_0/\mu\sqrt{2}$ at low $T$. 
Observation of the change of the transition type with $x$ 
would confirm unambiguously existence of the LOFF state at $x=0$. 
Interestingly, as $x$ tends to unity, the system becomes clean again, and the 
LOFF state may reappear. In fact, at $x=1$ the value of $H_{c2}\approx 14$ T 
at low $T$ is close to that anticipated for the 2D LOFF state. 
We note that $T_c$ increases slightly from 4.3 K at $x=0$ to 5.5 K at 
$x=1$.  Additionally, anisotropy probably drops with $x$, but even 
at $x=1$ the system remains close to 2D, because the interlayer correlation 
length is about the same as the interlayer spacing \cite{Ga}.  
Hence, both effects do not change our qualitative predictions.  

In conclusion, in $\lambda-({\rm BETS})_2{\rm FeCl}_4$ we anticipate 
first order phase transitions between the superconducting phases
corresponding to different higher Landau level states and 
between lattices with different 
symmetry.  We anticipate also the change of the S-N transition 
from second order to first order as $x$ approaches 1/2 in the system 
$\lambda-({\rm BETS})_2({\rm FeCl}_4)_{1-x}({\rm GaCl}_4)_x$.  
These phase transitions may be detected by magnetization 
and specific heat measurements.  
We think that neutron scattering, 
$\mu$SR, and NMR  measurements may be used additionally to reveal the 
structure of peculiar vortex phases 
formed due to the interplay of competing spin and orbital effects induced by 
the external and exchange fields.  

The authors, M.H. and A.B., are grateful to J.-P. Brison for useful 
discussions.  This work was supported by the Los Alamos National
Laboratory under the auspices of the US DOE and the ESF ``Vortex'' Program.


\begin{references}
\bibitem{Uji01}  S. Uji, {\it et al}, Nature {\bf 410}, 908
(2001).

\bibitem{Balicas01}  L. Balicas, {\it et al}, Phys. Rev. Lett. {\bf 87}, 7002 
(2001).

\bibitem{Jaccarino}  V. Jaccarino and M. Peter, Phys. Rev. Lett. {\bf 9},
290 (1962).

\bibitem{Meul84}H. W. Meul, {\it et al}, Phys. Rev. Lett. {\bf 53}, 497 (1984).

\bibitem{Lin85}  C. L. Lin, {\it et al}, Phys. Rev. Lett. {\bf 54}, 2541 
(1985).

\bibitem{Bulaevskii73}  L. N. Bulaevskii, Sov. Phys. JETP {\bf 37}, 1133
(1973).

\bibitem{FFLO}  P. Fulde and R. A. Ferrell, Phys. Rev. {\bf A135}, 550
(1964); A. I. Larkin and Yu. N. Ovchinnikov, Sov. Phys. JETP {\bf 20}, 762
(1965).

\bibitem{Bulaevskii74}  L. N. Bulaevskii, Sov. Phys. JETP {\bf 38,} 634
(1974).

\bibitem{Buzdin96}  A. I. Buzdin and J. P. Brison, Europhys. Lett. {\bf 35,}
707 (1996).

\bibitem{Helfand} E. Helfand and N. R. Werthamer, Phys. Rev. {\bf 147, }288
(1966).

\bibitem{Dupuis}A.I. Buzdin and S.V. Polonskii, Sov. Phys. JETP {\bf 66}, 422 
(1987 [Zh. Eksp. Teor. Fiz. {\bf 93}, 747 (1987)]; 
N. Dupuis, Phys. Rev. B, {\bf 51}, 9076 (1995). 

\bibitem{Abrikosov 57}  A. A. Abrikosov, Sov. Phys. JETP {\bf 5,} 1174
(1957).

\bibitem{Houzet 00}  M. Houzet and A. Buzdin, Europhys. Lett. 
{\bf 50}, 375 (2000).

\bibitem{Klein 00}  U. Klein, {\it et al}, J. Low Temp.
Phys. {\bf 118}, 91 (2000).

\bibitem{Buz-Kach}  A. I. Buzdin and H. Kachkachi, Phys. Lett. A {\bf 225},
 341 (1997).

\bibitem{Ga}M.A. Tanatar, {\it et al}, J. Superc. {\bf 12}, 511 (1999).
\end{references}
\end{document}